\documentclass[preprint,12pt]{elsarticle}
\usepackage{amsmath}

\usepackage{amssymb}

\usepackage{enumitem}   
\usepackage{soul}

\newcounter{bla}

\journal{Computer Physics Communications}

\begin{document}

\begin{frontmatter}



\title{PASTA: Python Algorithms for Searching Transition stAtes}


\author[label1]{Sudipta Kundu}
\author[label2]{Satadeep Bhattacharjee}
\author[label2]{Seung-Cheol Lee\corref{author1}}
\author[label1]{Manish Jain\corref{author2}}


\cortext[author1] {Corresponding author.\\\textit{E-mail address:} seungcheol.lee@ikst.res.in}
\cortext[author2] {Corresponding author.\\\textit{E-mail address:} mjain@iisc.ac.in}
\address[label1]{Centre for Condensed Matter Theory, Department of Physics, Indian Institute of Science, Bangalore 560012, India}
\address[label2]{Indo-Korea Science and Techonology Center, Bangalore 560065, India}

\begin{abstract}
Chemical reactions are often associated with an energy barrier along the reaction pathway
which hinders the spontaneity of the reaction.
Changing the energy barrier along the reaction pathway allows one to modulate the performance of a reaction.
We present a module, Python Algorithms for Searching Transition stAtes (PASTA), to calculate the energy barrier and locate the transition state of a reaction efficiently.
The module is written in python and can perform nudged elastic band, climbing image nudged elastic band and automated nudged elastic band calculations.
These methods require the knowledge of the potential energy surface (and its gradient along some direction). This
module is written such that it works in conjunction with density functional theory (DFT) codes to obtain this information.
Presently it is interfaced with three well known DFT packages: Vienna Ab initio Simulation
Package (VASP), Quantum Espresso and Spanish Initiative for Electronic Simulations with Thousands of
Atoms (SIESTA). This module is 
easily extendable and can be interfaced with other DFT, force-field or
empirical potential based codes.
The uniqueness of the module lies in its user-friendliness. For users with limited computing
resources, this module will be an effective tool as it allows to perform the calculations image by image.
On the other hand, users with plentiful computing resources (such as users in a high performance computing environment) can perform the calculations for large
number of images simultaneously. 
This module gives users complete flexibility, thereby enabling them to perform
calculations on large systems making the best use of the available resources.
\end{abstract}

\begin{keyword}
NEB \sep CI-NEB \sep AutoNEB \sep Transition State \sep Energy Barrier \sep python \sep DFT
\end{keyword}

\end{frontmatter}



{\bf PROGRAM SUMMARY}

\begin{small}
\noindent
{\em Manuscript Title:} PASTA: Python Algorithms for Searching Transition stAtes  \\
{\em Authors:}   Sudipta Kundu, Satadeep Bhattacharjee, Seung-Cheol Lee and Manish Jain                  \\
{\em Program Title:}    PASTA       \\
{\em Program obtainable from:} \verb|https://www.ikst.res.in/pasta.php| \\
{\em Journal Reference:}                                      \\
{\em Catalogue identifier:}                                   \\
{\em Licensing provisions:} Open source BSD License                \\
{\em Programming language:}  Python                              \\
{\em Computer:} Any computer with Python installed. The code has been tested with Python2.7 and Python3.5.       \\
{\em Operating system:} Unix/Linux                    \\
{\em Keywords:} NEB, CI-NEB, AutoNEB, Transition State, Energy Barrier, DFT\\
{\em External routines/libraries:} numpy, matplotlib \\
{\em Nature of problem:} Most of the reactions have an energy barrier on their reaction pathway.
This energy barrier affects the progress of the reaction.
{\em Solution method:} We implement the NEB, CI-NEB and AutoNEB method to locate transition state
and estimate the energy barrier. 
Our module works with density functional theory codes: VASP, SIESTA and Quantum Espresso presently.\\
\end{small}

\section{Introduction}

A large number of reactions involve movement of a group of atoms on potential
energy surface (PES) from one minimum to another. These minima can be
recognised as reactant (initial) and product (final) state of the reaction.
The reaction pathway is the minimum energy pathway (MEP) in the potential
energy surface (PES) that connects the reactants and products, along which the
reaction is most likely going to proceed. The maximum energy along the MEP
influences the rate of the reaction and associates an energy barrier with the
reaction.  For some reactions the barrier is comparable to room temperature. As
a result, such reactions are thermally activated. Other reactions may require
external energy to overcome the barrier. Knowledge of the energy barrier
and the reaction pathway are of paramount importance to regulate the
reaction or to design new chemical processes. 

The maxima along the MEP are saddle points on the PES. In general, more than one
saddle point can be present along the MEP. These saddle points correspond to
intermediate stable configurations. As a result, it is required to find the
saddle point with the highest energy to have a correct estimate of the
activation energy of the reaction.

There exist several well known methods of determining the reaction pathway:
nudged elastic band (NEB) \cite{neb1,neb2,neb5}, Lanczos based methods
\cite{lanczos}, activation-relaxation method \cite{activation} or the dimer
method \cite{dimer}. The latter two methods have the advantage that they do not
require the knowledge of the final products. However, in a large number of
applications, the reactants and products are known. In such cases, the NEB has
emerged as the leading method for determining the MEP.

The NEB \cite{neb1,neb2,neb5} method works by converging a trial initial path in the
PES to the MEP in the vicinity of the initial path. The resolution of the path
is defined by the number of images used to construct the path.
There also exists climbing image NEB (CI-NEB) which employs mostly the basic NEB method
but forces one of the images to be at the maximum of PES.
The NEB or CI-NEB method
needs the information of PES to find the MEP. For chemical reactions and
diffusions, density functional theory (DFT) is the method of choice for
determining the PES. For the NEB/CI-NEB, the DFT calculation only needs to provide the
total energy and forces acting on the system of atoms. These quantities are
easily accessible within any DFT code such as Vienna Ab initio Simulation
Package (VASP) \cite{vasp1,vasp2,vasp3,vasp4,vasp5}, Quantum Espresso
\cite{qe}, Spanish Initiative for Electronic Simulations with Thousands of
Atoms (SIESTA) code \cite{siesta1,siesta2} etc. 
Furthermore, recently an automated procedure \textquoteleft
AutoNEB\textquoteright{} \cite{autoneb} has been proposed to efficiently locate
transition states. This method tries to use fewer resources than the NEB by
first converging a rough path before improving the resolution.

In this manuscript, we present a general implementation of NEB, CI-NEB and
AutoNEB  module which can be easily interfaced with any DFT code. The module 
Python Algorithms for Searching Transition stAtes (PASTA) is
written in python \cite{python} since the NEB, CI-NEB or AutoNEB method is not 
computationally intensive. Presently, it is interfaced with three different
DFT codes: VASP, SIESTA and Quantum Espresso.
We note that some of the codes already have an implementation of NEB.
Quantum Espresso \cite{qe} has the PWneb package and  Vasp Transition State
Theory Tools (VTST Tools) is available with VASP \cite{vasp1,vasp2,vasp3,vasp4,vasp5}.
Both these packages perform NEB and CI-NEB calculations and
are tightly bound to their respective DFT codes.
A recent implementation of NEB, CI-NEB as well as AutoNEB is a part of a larger
package named Atomic Simulation Environment (ASE) \cite{ase}.
Despite these implementations, our implementation is different in that
it not only provides the whole suite of NEB methods (including AutoNEB)
as a single package,
but also gives the user full control to run the DFT calculations.
While ASE launches the DFT calculations internally, the philosophy of this
module is to give the user maximum flexibility in terms of performing the
calculations. PASTA allows the user to run as many DFT calculations 
simultaneously as the user wants.
This ensures the usage of maximum resources available making the module
appropriate to use in high performance computing (HPC) system.
PASTA is also user friendly in the way that
it requires minimal effort to setup the input files to do the calculations.
Further more, the module is organized such that each step of NEB or CI-NEB or
AutoNEB calculation is performed by different routines permitting the user to
modify any part of the code with ease.
This feature also allows to add more interfaces to other DFT codes as well as
force field and empirical potential based codes.

We have tested PASTA for four test cases
which are described later.  This paper is organized as follows: section 2
describes the theory of NEB, CI-NEB and AutoNEB; work flow of these methods are described in section 3;
the structure of code and the input files are illustrated in section 4 and 5 respectively;
test systems are discussed in section 6 and we conclude in section 7.

\section{Theoretical framework}

\subsection{NEB}

Nudged elastic band method (NEB) is a chain-of-states method in
which several images (states) of the system are connected together to trace out
the path connecting the initial and final states.
The NEB method \cite{neb1,neb3,neb4} works by guessing an initial path
and then converging the \textquotedblleft trial\textquotedblright{} path
in the PES to the MEP in the vicinity of the initial path.
In this method, one starts by defining a number of images between the initial
and final states. The images may be interpolated along the straight line in the
PES connecting the two end points.
Each image is connected to adjacent images by Hooke's springs forming an elastic
band. The images are then moved simultaneously on the PES according to a force
projection scheme. The total force acting on the image consists of the
perpendicular component of the true force arising from the PES and the
parallel component arising from the spring force.
This force projection is known as \textquotedblleft nudging\textquotedblright{}
and this makes NEB different from other methods \cite{neb3,neb4}.
The perpendicular component of spring force influences the path to cut corner
where the MEP is curved and thus avoids the true saddle point.
On the other hand, the parallel component of true force forces the images to slide down
reducing the resolution around saddle point.
By using nudged force, the
\textquotedblleft corner-cutting\textquotedblright{}
and \textquotedblleft sliding-down\textquotedblright{} problems of elastic bands
are circumvented.
In this way, the spring force does not affect the convergence of the elastic
band to the MEP and the true force does not affect the distribution of images
along the MEP.

We have implemented the NEB module in PASTA as described by Henkelman et. al. in ref.\cite{neb1}.
Let us consider an elastic band of N+1 images and denote the images by [\textbf{R$_0$},
\textbf{R$_1$},\textbf{R$_2$},\ldots,\textbf{R$_N$}] where
\textbf{R$_0$} and \textbf{R$_N$} correspond to initial and
final state respectively.
Given an estimate of the tangent ($\boldsymbol{\hat{\tau}_{i}}$) to the path at image i,
the total force acting on image i is given by:
\begin{equation}
\label{eqn1}
\textbf{F$_i$} = -\nabla \mathrm{V}(\textbf{R$_i$})\lvert_\perp + \textbf{F$_{i}^{s}$}\lvert_\parallel
\end{equation}
where the first term represents the perpendicular component of true force arising from PES and is given by:
\begin{equation}
\label{eqn2}
\nabla \mathrm{V}(\textbf{R$_i$})\lvert_\perp = \nabla \mathrm{V}(\textbf{R$_i$}) - \nabla \mathrm{V}(\textbf{R$_i$}).\boldsymbol{\hat{\tau}_{i}}\boldsymbol{\hat{\tau}_{i}}
\end{equation}
The second term is the parallel component of spring force which is evaluated as below \cite{neb1}:
\begin{equation}
\label{eqn3}
\textbf{F$_{i}^{s}$}\lvert_\parallel = \frac{1}{2}((k_{i+1}+k_{i})\lvert \textbf{R$_{i+1}$} - \textbf{R$_{i}$}\lvert - (k_{i}+k_{i-1})\lvert \textbf{R$_{i}$} - \textbf{R$_{i-1}$}\lvert)\boldsymbol{\hat{\tau}_{i}}
\end{equation}
If the spring constant is same for all images, the expression reduces to the following:

\begin{equation}
\label{eqn4}
\textbf{F$_{i}^{s}$}\lvert_\parallel = k(\lvert \textbf{R$_{i+1}$} - \textbf{R$_{i}$}\lvert - \lvert \textbf{R$_{i}$} - \textbf{R$_{i-1}$}\lvert)\boldsymbol{\hat{\tau}_{i}}
\end{equation}
The local tangent is calculated as follows:
\begin{equation}
\label{eqn5}
    \boldsymbol{{\tau_{i}}}=
    \begin{cases}
     \boldsymbol{\tau_{i}^{+}} , & \text{if}\ V_{i+1} > V_{i} > V_{i-1}\\
     \boldsymbol{\tau_{i}^{-}} , & \text{if}\ V_{i+1} < V_{i} < V_{i-1}
    \end{cases}
\end{equation}
where,
\begin{equation}
\label{eqn6}
\boldsymbol{\tau_{i}^{+}} = \textbf{R$_{i+1}$} - \textbf{R$_{i}$} \hspace{1cm}  \text{and} \hspace{1cm} \boldsymbol{\tau_{i}^{-}} = \textbf{R$_{i}$} - \textbf{R$_{i-1}$}
\end{equation}
and $V_{i}$ is the energy at image i.

If image i is at a minima (V$_{i+1}$$>$V$_{i}$$<$V$_{i-1}$) or a maxima (V$_{i+1}$$<$V$_{i}$$>$V$_{i-1}$), then the tangent becomes,

\begin{equation}
\label{eqn7}
    \boldsymbol{{\tau_{i}}}=
    \begin{cases}
     \boldsymbol{\tau_{i}^{+}}\mathrm{\Delta}V_i^{max}+\boldsymbol{\tau_{i}^{-}}\mathrm{\Delta}V_i^{min} , & \text{if}\ V_{i+1} > V_{i-1}\\
     \boldsymbol{\tau_{i}^{+}}\mathrm{\Delta}V_i^{min}+\boldsymbol{\tau_{i}^{-}}\mathrm{\Delta}V_i^{max} , & \text{if}\ V_{i+1} < V_{i-1}
    \end{cases}
\end{equation}
where,
\begin{equation}
\label{eqn8}
\mathrm{\Delta}V_i^{max} = max(\lvert V_{i+1}-V_{i}\lvert , \lvert V_{i-1}-V_{i}) \hspace{.3cm} \text{and} \hspace{.3cm} \mathrm{\Delta}V_i^{min} = min(\lvert V_{i+1}-V_{i}\lvert , \lvert V_{i-1}-V_{i})
\end{equation}
With this prescription, the images between the two fixed end points are moved
according to the force given by eqn.(1) to obtain the MEP. 

\subsection{Interpolation}
In many cases, none of the images represent exactly the saddle point
even after the convergence has been achieved. So, interpolation between images
is required to find the saddle point energy. A cubic polynomial is used
to represent the MEP between each pair of adjacent images \cite{neb1}.
The polynomial for the segment [\textbf{R$_i$},\textbf{R$_{i+1}$}] is written as: $a_{i}x^3 + b_{i}x^2 + c_{i}x + d_{i}$ .
The parameters are determined using the information of the energies and
forces at the end points of the segment and they are given as below:
\begin{eqnarray}
\label{eqn9}
a_{i} = \frac{2(V_{i}-V_{i+1})}{R_{i}^3} - \frac{F_i+F_{i+1}}{R_{i}^2}
\label{eqn10}
\\b_{i} = \frac{3(V_{i+1}-V_{i})}{R_{i}^2} + \frac{2F_i+F_{i+1}}{R_{i}}
\label{eqn11}
\\c_{i} = -F_{i}\\
\label{eqn12}
d_{i} = V_{i}
\end{eqnarray}
Here, $R_{i}$ is the length of the ith segment and $V_{i}$
and $V_{i+1}$ are the energies at two end points and $F_{i}$
and $F_{i+1}$ are the parallel forces of them.

\subsection{Climbing Image NEB}
Climbing image NEB (CI-NEB) \cite{neb2} is an improvement over the regular NEB method in
finding the transition state. In this method, the image with the highest energy
after few regular NEB steps, is made to move uphill in energy along the elastic
band. This is accomplished by removing the spring force on this image and
including the inverted parallel component of true force.
The force acting on this image is given by:
\begin{equation}
\label{eqn13}
\textbf{F$_{i_{max}}$} = -\nabla \mathrm{V}(\textbf{R$_{i_{max}}$}) + 2\nabla \mathrm{V}(\textbf{R$_{i_{max}}$}).\boldsymbol{\hat{\tau}_{i_{max}}\hat{\tau}_{i_{max}}}
\end{equation}
where, \textbf{R$_{i_{max}}$} corresponds to the image with the highest energy
and $\boldsymbol{\hat{\tau}_{i_{max}}}$ is the tangent at that image.

In CI-NEB method, the climbing image converges to the saddle point providing
the correct saddle point energy within tolerance.

\subsection{AutoNEB}
Automated nudged elastic band (AutoNEB) \cite{autoneb} uses the basic NEB algorithm.
It divides the whole process in two subprocesses: (i) initializing the path
and (ii) converging the path.
The strategy of AutoNEB is to handle a subset of total images ($N_{sim}$ of $N+1$ images)
at a time and to converge the path roughly.
To initialize the path, normal NEB is started with a given number of images ($N_{sim}$)
and the path is roughly converged.
The path is declared roughly converged if the maximum force on any of the images
goes below the given threshold or if the allowed number of force evaluation
steps per image is reached. Next a new image is added and the piece of path centering the newly added image (with $N_{sim}$ images) is again converged.
The new image is added either in the biggest geometrical gap or energy gap
following the inequality (eqn. 14). If the inequality is satisfied an image is
inserted in the biggest geometrical gap, else the image is added in the largest
energy gap. 

\begin{eqnarray}
\label{eqn14}
\frac{f_{1}(\{\textbf{R$_i$}\})}{f_{2}(\{E_i\})} > r_{se}
\label{eqn15}
\\f_{1}(\{\textbf{R$_i$}\}) = \frac{max(|\textbf{R$_{i+1}$ - R$_i$}|)}{|\textbf{R$_N$ - R$_0$}|}
\label{eqn16}
\\f_{2}(\{E_i\}) = \frac{max(\Delta E'_i)}{E_{max} - E_{min}}
\end{eqnarray}
Here $r_{se}$ is a ratio specified by user and $\Delta E'_i$ is scaled energy difference.
\begin{eqnarray}
\label{eqn17}
\Delta E'_i = \Delta E_i \frac{E_{avg,i}}{E_{max} - E_{min}}\\
\label{eqn18}
\Delta E_i = |E_{i+1} - E_i|\\
\label{eqn19}
E_{avg,i} = \frac{E_{i+1}  + E_i - 2E_{min}}{2}
\end{eqnarray}
Once a new image is added, spring constants connected to that image have to be
updated. This addition of more images is continued until maximum number of
images or the energy and geometrical resolution is achieved.
Then the rough path containing N + 1 images is converged with CI-NEB scheme.

\section{Workflow}

The general steps involved in NEB/CI-NEB method to calculate energy barrier of
a reaction (fig. 1(a)) are followings:

\begin{enumerate}

\item A trial initial path is constructed by interpolating between initial and final images.
\item DFT calculations are performed with the images to obtain energies and true forces (originating due to PES)
acting on the images.
\item The nudged forces are calculated according to NEB/CI-NEB scheme
and the images are moved on PES.
The updated postions are passed to DFT codes if the images are not converged.

\end{enumerate}

Step 2 and 3 are repeated until nudged forces on all images are converged.

\begin{enumerate}[resume]
\item Converged images are interpolated to get the energy barrier.
\end{enumerate}

Step 1, 3 and 4 are performed by PASTA and
step 2 is done using DFT code of user's choice (among Quantum Espresso, VASP and SIESTA).

\begin{figure}[h]
\begin{center}
\includegraphics[scale=0.45]{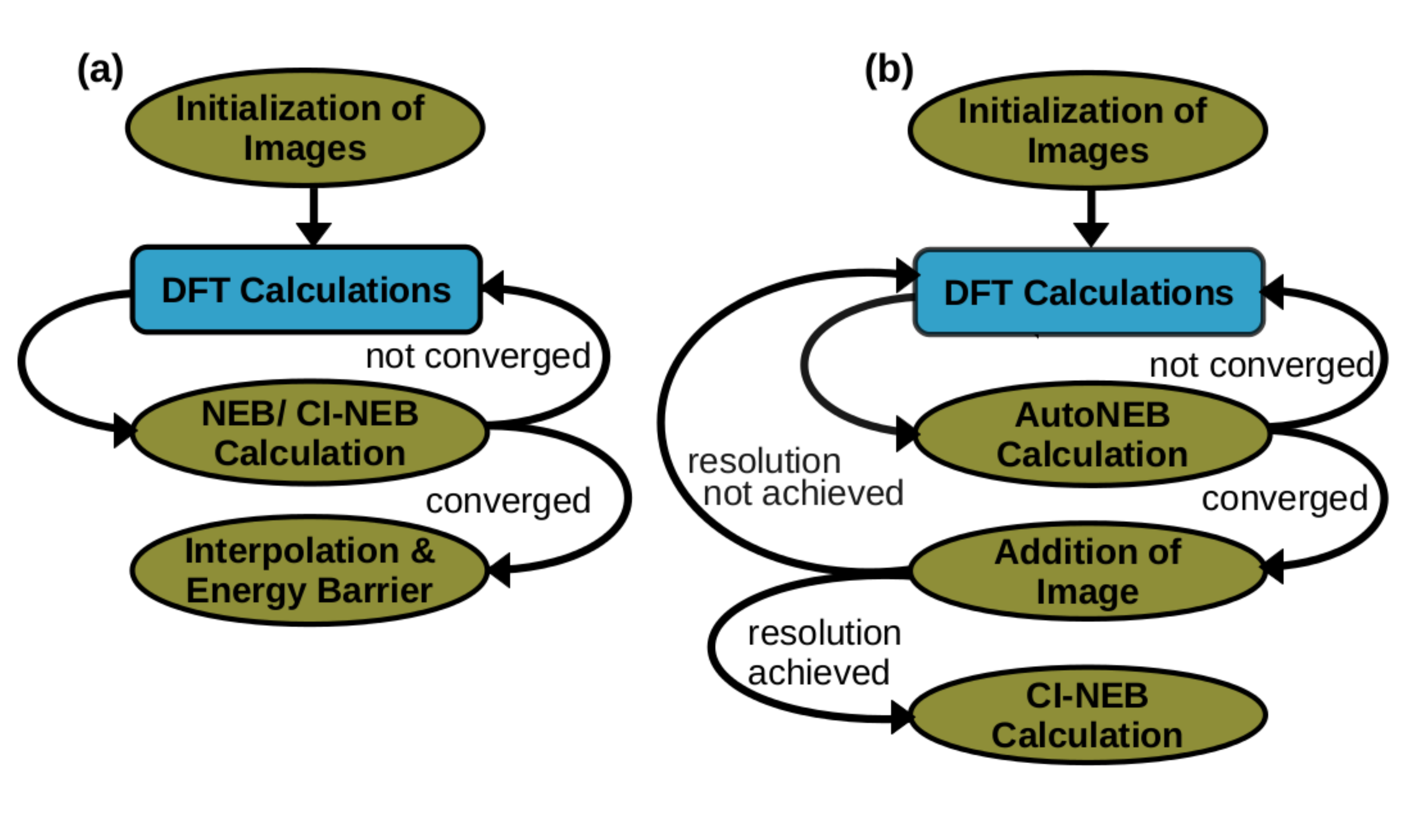}
\caption{\label{fig.1}(a) and (b) describes the work flow of NEB/CI-NEB and AutoNEB respectively. The olive ellipses are done with PASTA and the blue rectangle denotes the DFT calculations.}
\end{center}
\end{figure}

AutoNEB is different from NEB/CI-NEB in constructing the initial path.
While constructing the path, AutoNEB deals with a piece of path.
The steps of an AutoNEB calculation (fig. 1(b)) are illustrated below:
\begin{enumerate}
\item The trial path is initialized with a fewer number of images.
\item DFT calculations have to be performed with those images.
\item Images are moved according to nudged forces and the updated images are
passed to DFT code.
\end{enumerate} 
Step 2 and 3 are continued until the piece of path is converged.
\begin{enumerate}[resume]
\item After the piece of path is converged, a new image is added in the path.
A different piece of path centering the new image is selected.
\end{enumerate}
Now step 2 and 3 are again continued with the selected path to add another image.
In this way step 2, 3 and 4 are performed to sample the path well enough.
\begin{enumerate}[resume]
\item When the resolution is achieved, CI-NEB calculation is performed with the full path to obtain the energy barrier.
\end{enumerate}
CI-NEB has been described before.
Although in this case initialization of path is not required
since the path is constructed by AutoNEB.

\section{Code Layout}
We have implemented the NEB, CI-NEB and AutoNEB module as described in \cite{neb1,neb2,autoneb}.
The PASTA code is written in python. It consists of the primary routines
for NEB and AutoNEB separately, the interface with DFT codes and the optimizer.
The code requires 3 input files: 2 separate input files with initial and final
coordinates and a \textquoteleft neb\_params\textquoteright{} file containing
the parameters required for NEB, CI-NEB and AutoNEB calculation, which is described in
next section. It also requires all the files (like pseudopotential files) needed for DFT calculation.
Sample input files are provided in 
\textquoteleft Example\textquoteright{} folder of the PASTA code.

Upon unzipping the tar file, the main folder holds two sub-folders.
The \textquoteleft Code\textquoteright{} folder contains 5 sub-folders:
NEB, AutoNEB, Interface, Optimizer and Utility and a file \textit{pasta.py}.
It also contains scripts to setup the code.
\textit{pasta.py} is the main executable and calls the appropriate
routines required to run NEB, CI-NEB and AutoNEB as specified by the user.
In the following subsections, the functions of routines are described. 

\subsection{NEB}
This folder contains the routines required to run NEB or CI-NEB. The primary
routines are the followings:

(a)\textit{initialize.py}: This routine interpolates between two end point
images along the straight line connecting them and put each of the images in
separate folders. It will create \textquoteleft pos.save\textquoteright{} and
\textquoteleft folder.save\textquoteright{} files which hold the positions of
all images and the list of folders respectively in the pickle data format.

(b) \textit{pot\_frc.py} : After the inputs in every folder  are run with DFT
codes, this routine parses the output files to obtain the energy and the true
force acting on each image. It writes out the energies and forces in
\textquoteleft V.save\textquoteright{} and
\textquoteleft F.save\textquoteright{} files respectively. 

(c) \textit{neb.py} : neb.py calculates the force acting on each image using
eqn.(1) and writes them in \textquoteleft 
force\_neb.save\textquoteright{}.
It also reports the path converged when the nudged forces on all images are less
than the threshold provided by the user.

(d) \textit{neb\_climb.py} : If climbing image NEB is to be used, this routine
is used instead of neb.py. It uses eqn.(13) to calculate the force for the image
with the highest energy and eqn.(1) for the rest of the images.

(e) \textit{pos\_updt.py} : This routine updates the position of the moving
images at each iteration using the optimizer specified by the user.

\begin{figure}[h]
\begin{center}
\includegraphics[scale=0.4]{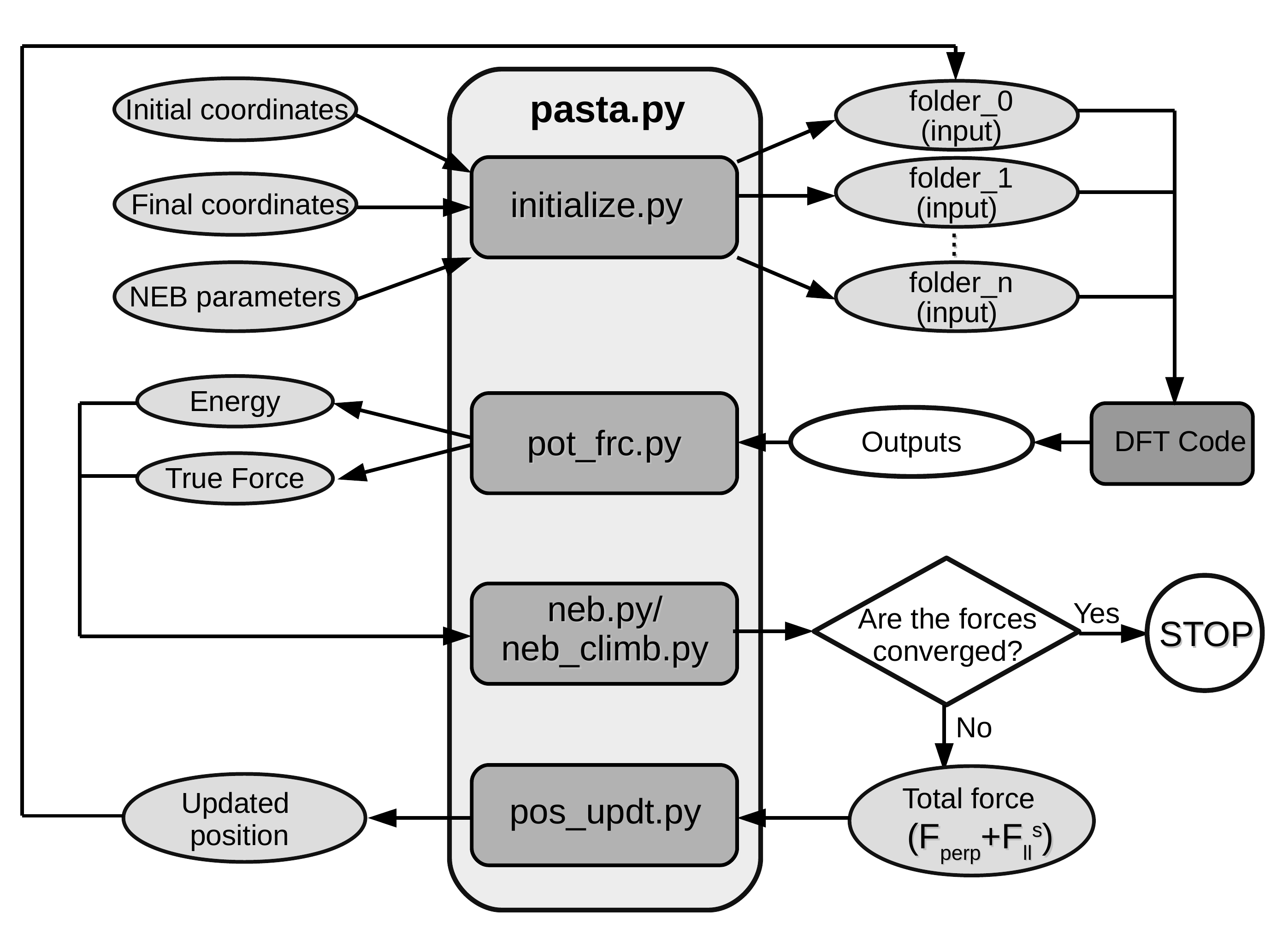}
\caption{\label{fig.2} Flowchart of the NEB/CI-NEB code}
\end{center}
\end{figure}

To execute the NEB/CI-NEB code, user has to run the main script
\textit{pasta.py} to initialize the folders. After running DFT
calculations in the set of folders, again \textit{pasta.py} has to be
run. This process should continue until the force on all the moving images go
down below the given tolerance. As the end points are kept fixed, user may run
DFT calculation in the first and last folders once. The working process of NEB/CI-NEB
code is illustrated in fig. \ref{fig.2}. The code writes out  the maximum
component of force on each moving image at every iteration in the file
\textquoteleft PASTA.out\textquoteright{} and details of the calculation in
\textquoteleft PASTA.log \textquoteright{}. Once the convergence has
been achieved, the code stops after generating a file
\textquoteleft Converged\textquoteright.


\subsection{AutoNEB}
The AutoNEB folder contains the following routines:

(a) \textit{initialize.py}: This routine is same as the routine in NEB except
that it writes out \textquoteleft pos\_tot.save\textquoteright{} and
\textquoteleft folder\_tot.save\textquoteright{} instead of
\textquoteleft pos.save\textquoteright{} and
\textquoteleft folder.save\textquoteright{} respectively. It also initializes
the array of spring constants in \textquoteleft k\_tot.save\textquoteright{}
file. AutoNEB requires this because the spring constants between images have to
be updated after a new image is added.

(b) \textit{pot\_frc.py}: This  routine performs the same job as in NEB. In
AutoNEB a subset of total images ($N_{sim}$ of $N+1$ images) has to be run at a
time. So, this routine  calls \textit{reinitiate.py} to select those folders
($N_{sim}$ images) and create a list of them which is written in
\textquoteleft folder.save\textquoteright. It also writes out
\textquoteleft pos.save\textquoteright{} and
\textquoteleft k.save\textquoteright{} containing the positions and spring
constants of selected images. During first iteration, this routine also
initializes \textquoteleft V\_tot.save\textquoteright.

(c) \textit{neb.py}: This calculates the force acting on each image using
eqn.(\ref{eqn1}) and writes them in
\textquoteleft force\_neb.save\textquoteright. The difference of this routine
with the routine in NEB folder is that it reports the path converged if the
force on any of the image is less than threshold. 

(d) \textit{pos\_updt.py}: It updates the position of the moving images at each
iteration until the path is roughly converged or the maximum number of force
evaluation steps after adding a new image is reached. If any of the above
conditions is satisfied, this routine checks if the given geometrical or energy
resolution, if specified, is achieved by calling \textit{check\_resolution.py}
routine. If the resolution or the maximum number of images is not achieved, the
routine calls \textit{insert\_image.py} to insert a new image in the path and
writes out the new folder list with which DFT has to be run now. It also updates
the \textquoteleft pos\_tot.save\textquoteright{},
\textquoteleft V\_tot.save\textquoteright{},
\textquoteleft k\_tot.save\textquoteright{} and
\textquoteleft folder\_tot.save\textquoteright{}. And if the resolution is
achieved, the routine instructs the user to start CI-NEB with all the folders.

\begin{figure}[h]
\begin{center}
\includegraphics[scale=0.4]{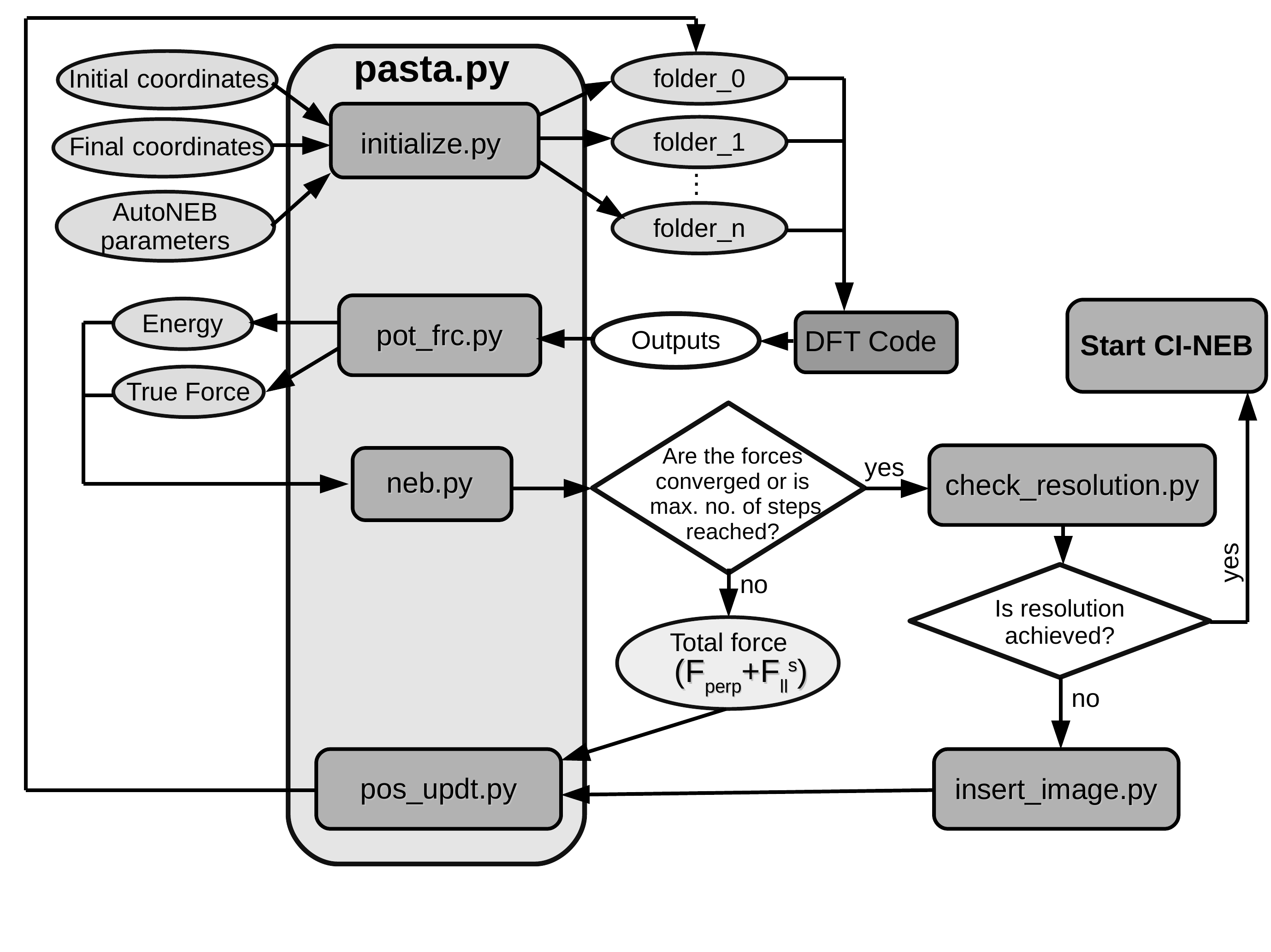}
\caption{\label{fig.3} Flowchart of the AutoNEB code}
\end{center}
\end{figure}

To execute the AutoNEB code, the \textit{pasta.py} and the DFT
calculations in the given set of folders have to be run iteratively one afte
another like NEB calculation (fig. \ref{fig.3}). But the difference is that if
the current piece of path is roughly converged while running,
\textquoteleft Converged\_temp\textquoteright{} will be generated by
\textit{neb.py} and a new image will be added by \textit{pos\_updt.py}. After
adding new image, the DFT code has to be run in a new set of folders. This
folder list will also be created by \textit{pos\_updt.py} in
\textquoteleft folder.save\textquoteright{} file. When the path is well sampled
user has to run CI-NEB with all the folders.

\subsection{Interface}
The interface folder contains all the routines required for read-write operations. 

\textit{read\_neb.py}  reads the parameters required for NEB or CI-NEB or AutoNEB
calculation from \textquoteleft neb\_params\textquoteright{} file. 

The code is interfaced with SIESTA, Quantum Espresso and VASP through
\textit{interface\_siesta.py}, \textit{interface\_qe.py} and
\textit{interface\_vasp.py} respectively. These routines read the initial and
final structure from the input files and generate the input files for the
interpolated images. They also extract the potential and force from the outputs
once DFT calculations are performed with the input files. This structure of the
code provides the flexibility of adding interfaces with other DFT, force-field
and empirical potential based codes also.
 
\subsection{Optimizer}
\textit{optimizer.py} in the Optimizer folder implements 4 optimization
algorithms \cite{optm}: steepest descent (SD), conjugate gradient (CG),
quick-min (QM) and Broyden-Fletcher-Goldfarb-Shanno (BFGS) to find the minima
in PES. The AutoNEB code only allows steepest decent and quick-min optimizers.

\subsection{Utility}
Utility contains some additional routines to analyze and visualize the results.

\textit{interp\_en.py} interpolates between images using eqn.(\ref{eqn9}-\ref{eqn12})
and plots energy as a function of reaction coordinate. It also gives the barrier
of the reaction.

\textit{write\_axsf.py} writes the positions and nudged forces (in Hartree unit) of all images in axsf format. This
routine can be called from the working directory at any point of calculation.
This will take the current structure and nudged force from all the folders and write them in
\textquoteleft file\_axsf\textquoteright. The file can be visualized with
Xcrysden \cite{xcrys}.

\section{Input Files}
The execution of PASTA module requires all the files needed to run the
DFT calculations; for example: the pseudopotential files for Quantum Espresso
and SIESTA and INCAR, POTCAR, KPOINTS for VASP. Further there should be two
separate files holding the coordinates for initial and final images in the
appropriate format specific to the DFT code. The file
\textquoteleft neb\_params\textquoteright{} has to be also provided. Sample
input files are given in the \textquoteleft Example\textquoteright{} folder. The
parameters for \textquoteleft neb\_params\textquoteright{} are described below:

\begin{verbatim}
DFT ESPRESSO         # Code for DFT calculation:
                     # ESPRESSO/SIESTA/VASP
NEB_METHOD AutoNEB   # Method of NEB: NEB/CI-NEB/AutoNEB
OPT QM               # Optimizer: SD/CG/QM/BFGS for NEB/CI-NEB
                     # SD/QM for AutoNEB
                     
## neb parameters
# These parameters are required by any of the 3 NEB methods
SpringConstant 5.0   # Spring constant in eV/angstrom^2
ForceThreshold 0.05  # Tolerance for force; in eV/angstrom
StepSize 0.2         # Step size in direction of force
MaxStep 0.3          # Maximum step allowed

FreezeMatrix         # Constrain the motion of atom
1 0 0                # Set to 0 to freeze the motion of atom 
0 0 0                # in a direction
1 0 0                # Set 1 otherwise

## NEB/CI-NEB
# Parameter required for only NEB and CI-NEB
Number of images 7   # integer number

## CI-NEB
# Parameter required for only CI-NEB
ClimbStep 2          # The image with the highest energy starts
                     # climbing uphill only after this many 
                     # number of steps have completed with NEB.
                     # If unspecified CI-NEB starts from 1st step
                     
## AutoNEB
# Parameters required for only AutoNEB
NumberOfStartingImages 5      # integer number
NumberOfSimultaneousImages 3  # Odd integer; less than number
                              # of images atleast by 2
NumberOfMaximumImages 15
ForceStepPerIteration 4       # Number of force evaluation steps
                              # per image after adding new image
Ratio_se 0.8                  # between 0 to 1
EnergyResolution 0.2          # Specify if energy resolution 
                              # is required; in eV
GeometricalResolution 0.5     # Specify if geometrical resolution 
                              # is required. 
#Both the resolution criteria can be specified. If none of them is
#specified, maximum number of images will be considered as stopping 
#criterion.
\end{verbatim}

\section{Test Systems}
We have calculated energy barrier for four different examples using this PASTA module. 
In all the cases, 
the force threshold is set to 0.05 eV/\AA{}.

\subsection{Example 1}


This example considers a simple reaction:
\begin{center}
H$_2$ + H $\longrightarrow$ H + H$_2$
\end{center}

\begin{figure}[]
 \begin{center}
 \includegraphics[scale=0.45]{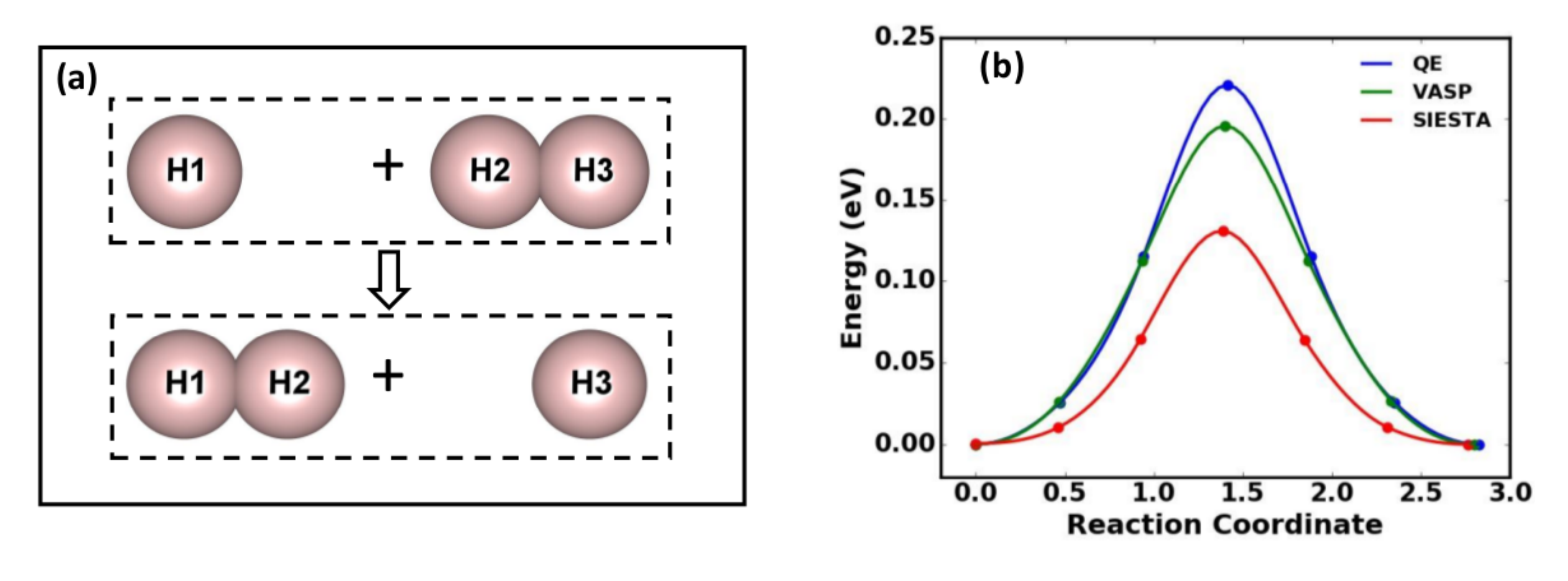}
 \caption{\label{fig.4}(a) shows the initial and final state of the reation. (b) is the interpolated energy vs. reaction coordinate plot for example 1 with three DFT codes}
 \end{center}
 \end{figure}

In this triatomic reaction involving hydrogen, 
one of the atoms rearranges itself by
bond dissociation with one and formation with another (fig. \ref{fig.4}(a)).
We calculate the energy barrier of this reaction.
This example was performed with all three DFT codes: Quantum Espresso (QE) \cite{qe}, VASP \cite{vasp1,vasp2,vasp3,vasp4,vasp5} and
SIESTA \cite{siesta1,siesta2}. We used norm-conserving pseudopotential for Quantum Espresso and SIESTA;
VASP calculations were performed with PAW potential \cite{paw1,paw2}. The exchange correlation
functional was approximated by GGA-PBE functional \cite{pbe}. DFT calculations
were done on gamma point only. The wave functions in Quantum Espresso and VASP
were expanded using plane waves upto energy of 50 Rydberg and 250 eV
respectively. Double zeta plus polarization orbitals were used as basis set in
SIESTA. The CI-NEB calculation was done with 7 images. The step length was chosen
0.2 and the spring constant was 5.0 eV/\AA{}$^2$.
Fig. \ref{fig.4}(b) shows the plot of interpolated energy versus
reaction coordinate using the CI-NEB method. The difference in the curves is due to
the differences in pseudo-potential and the basis sets used to run the DFT codes. The energy barrier calculated with Quantum Espresso, VASP and SIESTA are
221 meV, 196 meV and 131 meV respectively. We also calculated the energy barrier
using PWneb of Quantum Espresso for comparison and the energy barrier was found
to be 221 meV which is same as our result.
\begin{figure}[]
\begin{center}
\includegraphics[scale=0.45]{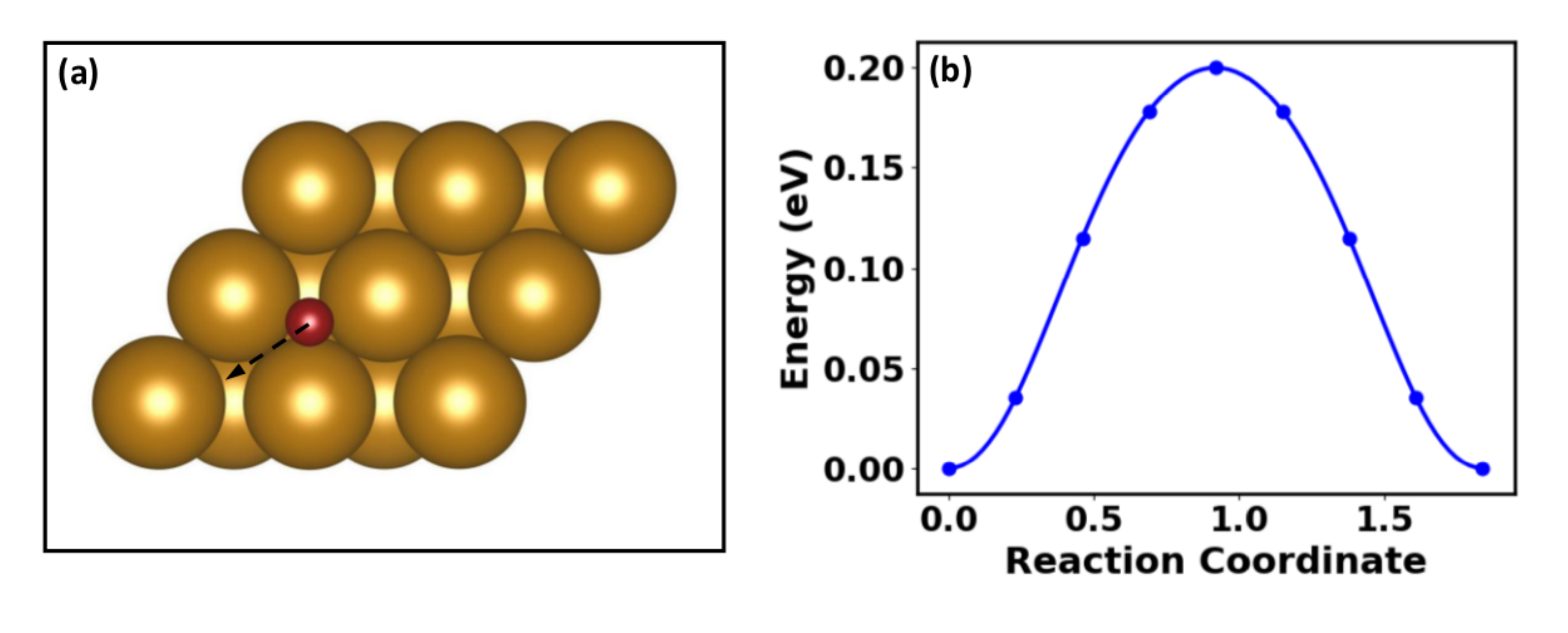}
\caption{\label{fig.5}(a) is the initial structure of H diffusion on Fe surface. Fe atoms are shown in golden color and the maroon one is H. The arrow shows where the H atom diffuses. (b) shows the interpolated energy vs. reaction coordinate plot}
\end{center}
\end{figure}

\subsection{Example 2 (Diffusion of hydrogen atom on iron surface)}

We calculate the energy barrier associated with the diffusion of hydrogen atom
on iron surface (BCC (110)). Iron is composed of BCC unit cell with lattice
constant 2.85\AA{} determined using pseudopotential mentioned below. A
2$\times$2 non-orthorhombic supercell with three layers was constructed to
simulate H diffusion on Fe surface. The bottom two layers were kept fixed to
mimic the bulk iron and the top layer along with the hydrogen atom were allowed
to move. The hydrogen finds its stable position at a threefold site and diffuses
from a threefold site (fig. \ref{fig.5}(a)) to another threefold site crossing
the short bridge site. To get an estimate of the energy barrier, we performed
CI-NEB calculation with 9 images. The step length was 0.1 and restricted to be
less than 0.15.
The spring constant was 5.0 eV/\AA{}$^2$.
The DFT calculations were performed using Quantum Espresso \cite{qe}. We used
ultrasoft psudopotential \cite{rrkjus} and the exchange-correlation functional was
approximated by revised Perdew-Burke-Ernzerhof generalized gradient functional
for solids and surfaces \cite{pbesol,sssp1,sssp2}. The energy
cutoff for expansion of wave functions (charge density) was chosen to be 100 Ry
(400 Ry). 8$\times$8$\times$1 k-point sampling was used. The CI-NEB calculation
gives the activation energy to be 200 meV (fig. \ref{fig.5}(b)) which is
slightly higher than the value calculated with RPBE \cite{fe_diifu} (180 meV).
In the transition state, the H atom resides on top of the short bridge site.


\subsection{Example 3 (Interstitial diffusion in silicon)}

\begin{figure}[h]
\begin{center}
\includegraphics[scale=0.45]{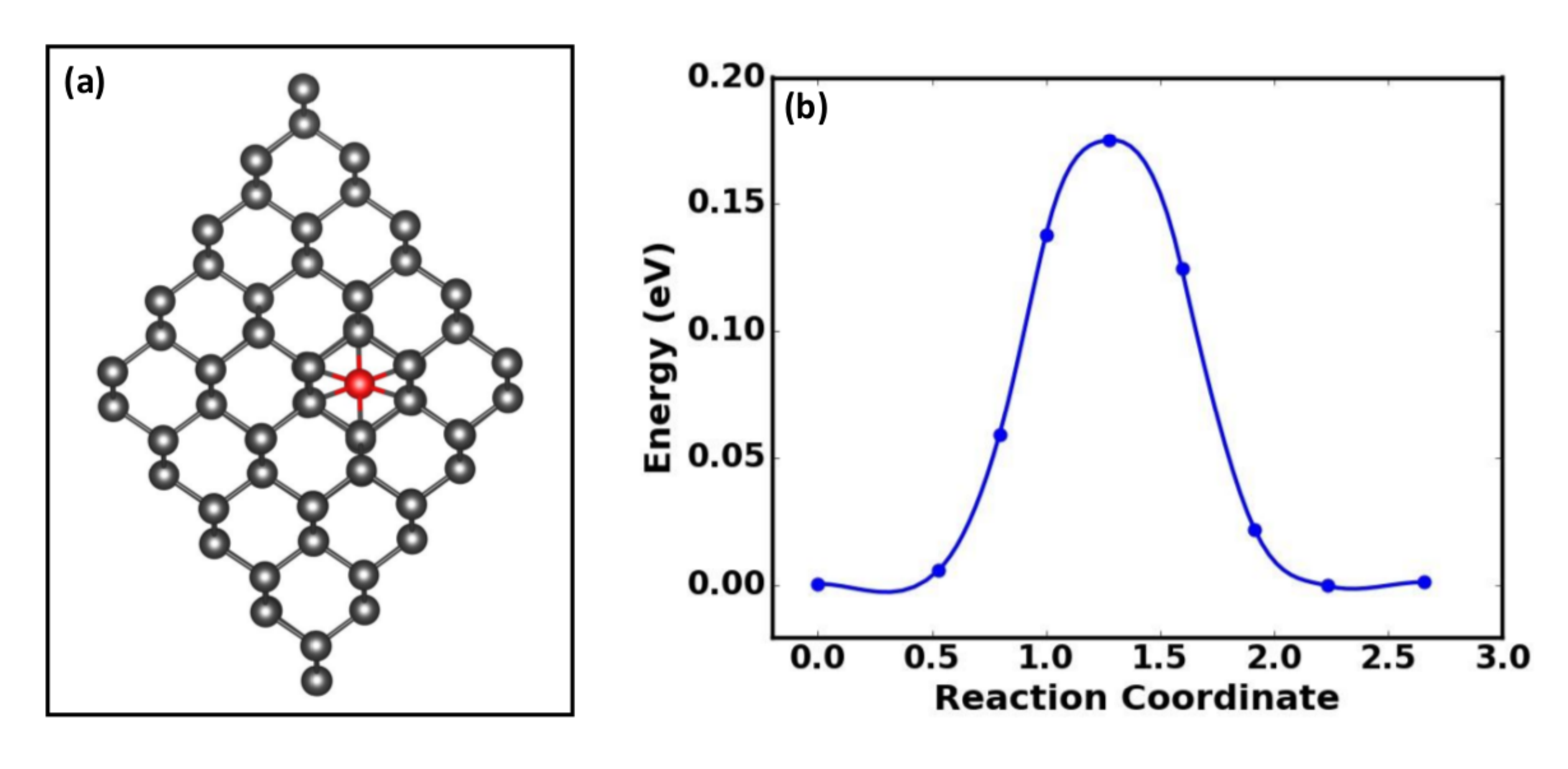}
\caption{\label{fig.6}(a) shows the self-interstitial in Si. The interstitial atom is red colored. (b) shows the interpolated energy vs. reaction coordinate plot. This plot is obtained using AutoNEB.}
\end{center}
\end{figure}

Now we investigate the interstitial diffusion in silicon. Many structures are
possible for self-interstitial in silicon \cite{si1,si2}. Hexagonal structure is
one of the low energy structures among them. We focus on the hopping of the
self-interstitial from one hexagonal site (fig. \ref{fig.6}(a)) to one of the
nearest hexagonal sites. To simulate the self-interstitial, we used a
4$\times$4$\times$4 supercell of cell dimension 21.82\AA{}.  This cell is large
enough to prevent interaction between interstitial atoms. Quantum Espresso
\cite{qe} was used for DFT calculations. Norm-conserving pseudopotential was
used to describe electron-ion interactions and the exchange-correlation was
approximated with GGA-PBE \cite{pbe} functional. The wave functions were
expanded using plane waves upto 30Ry energy and 2$\times$2$\times$2 k-grid was
used. To get the energy barrier, we used AutoNEB here. Spring constant was
chosen to be 5.0 eV/\AA{}$^2$.
We started with 5 images and the code added 4 more images to achieve the given
resolution. The energy resolution was specified 0.2 eV and the geometrical
resolution was 0.5. The step size was 0.2 initially but as the code kept adding
images, the step size decreased accordingly.  Once the roughly converged path
was obtained with 9 images, climbing image NEB was performed to get an energy
barrier of 175 meV (fig. \ref{fig.6}(b)) which is close to the value 180 meV as
found by R. J. Needs \cite{si2}. 

\subsection{Example 4 (CO dissociation on iron surface)}

\begin{figure}[]
\begin{center}
\includegraphics[scale=0.45]{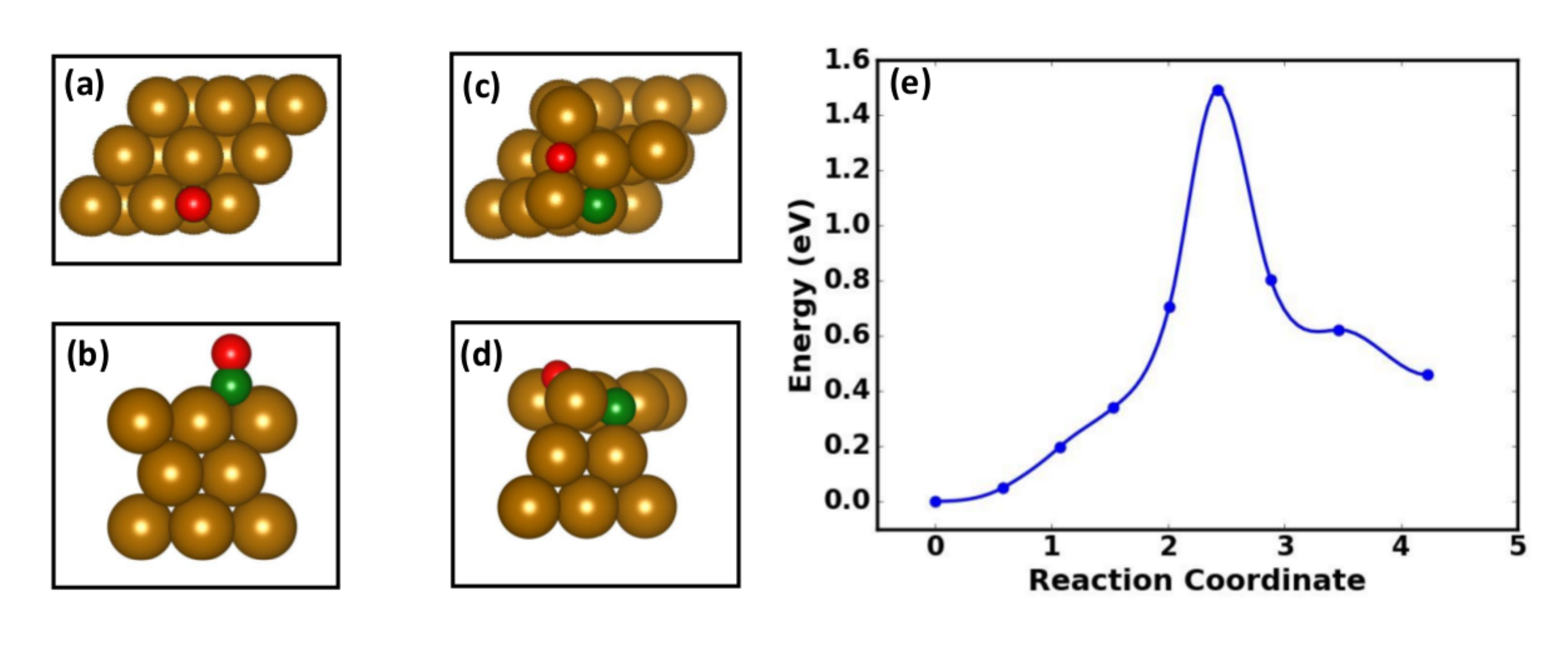}
\caption{\label{fig. 7}(a) and (b) shows the top view and side view of CO adsorbed on Fe surface respectively. Only 3 top layers of the supercell are shown in side view.  The golden atoms are Fe, C is colored green and O is red. (c) and (d) shows the dissociated CO on Fe surface. And (e) is the interpolated energy versus reaction coordinate plot.}
\end{center}
\end{figure} 

Next we investigate the dissociation of a carbon monoxide molecule on iron (BCC
(110)) surface. The DFT calculations was performed with VASP \cite{vasp1,vasp2,vasp3,vasp4,vasp5}. PAW potential was used \cite{paw1,paw2} and  exchange
correlation functional was approximated with GGA-PBE \cite{pbe} functional. We
used a 2$\times$2 7-layer supercell for the calculation. The unit cell lattice
constant was 2.83\AA{} \cite{co} and the energy cutoff for expanding wave
functions was 450 eV. 8$\times$8$\times$1 k-sampling was used to simulate the
surface. The bottom 4 layers were kept fixed during relaxation and CI-NEB calculation both. 
Although on top position of CO on Fe is a stable structure for CO adsorption on
iron surface, CO in long bridge site has a weaker bond. So, to study the
dissociation of CO, we started with CO at long bridge site (fig. \ref{fig. 7}(a) and \ref{fig. 7}(b)).
When CO molecule gets dissociated on iron surface, the bond between C and O
breaks and the C molecule gets adsorbed into the surface (fig. \ref{fig. 7}(c)
and \ref{fig. 7}(d)).
Now, we found the energy barrier of dissociation of CO using CI-NEB. The number
of images used was 9 and the step size wass 0.1. If we chose a larger step size,
some of the images kept moving back and forth. Fig. \ref{fig. 7}(e) depicts how
the energy varies with images.
The obtained energy barrier is 1.49 eV which agrees well
with literature\cite{co}.

\section{Conclusion}
We present a module PASTA to compute energy barrier and locate transition
state of a reaction. The module has been written in python and is interfaced
with Quantum Esresso, SIESTA and VASP presently. The module 
can be easily interfaced to other DFT codes, force field and empirical potential based codes as well.
We have tested the code in four cases: a bond formation and dissociation reaction involving hydrogen atoms,
diffusion of hydrogen atom on iron surface, interstitial hopping in silicon and
dissociation of CO on iron surface. The energy barriers we obtained are in good
agreement with literatures.

\section{Acknowledgments}
This work has been supported by Global Knowledge Perform (GKB) project through
Indo-Korea Science and Technology Center (IKST).
We thank the Supercomputer Education and Research Centre (SERC) at Indian Institute of Science (IISc) and Korea Institute of Science and Technology (KIST) 
for providing the computational facilities.







\section*{References}

\bibliographystyle{elsarticle-num}







\end{document}